\documentclass[reqno,a4paper,12pt]{article}
\usepackage{amsmath}
\usepackage{amssymb}
\usepackage[dvips]{epsfig}
\usepackage{graphicx}

\begin{document}

\begin{center}
{\Large\bf Narrowing the window 
for millicharged particles by CMB anisotropy}\\
\vspace{0.3cm}
S.~L.~Dubovsky$^a$\footnote{{\bf e-mail}: sergd@ms2.inr.ac.ru}, 
D.~S.~Gorbunov$^{a}$\footnote{{\bf e-mail}: gorby@ms2.inr.ac.ru},
G.~I.~Rubtsov$^{a,b}$\footnote{{\bf e-mail}: grisha@ms2.inr.ac.ru} 
\\
$^a${\small{\em 
Institute for Nuclear Research of the Russian Academy of Sciences, }}\\
{\small{\em
60th October Anniversary prospect 7a, Moscow 117312, Russia
}}\\
$^b${\small{\em 
Moscow State University, Department of Physics,}}\\
{\small{\em
Vorobjevy Gory, Moscow, 119899, Russia
}}
\end{center}

\begin{abstract}
We calculate the cosmic microwave background (CMB) anisotropy spectrum
in models with millicharged particles of electric charge $q\sim
10^{-6}-10^{-1}$ in units of electron charge. We find that a 
large region of the parameter space for the millicharged particles
exists where their effect
on the CMB spectrum is similar to the effect of
baryons. Using WMAP data on the CMB anisotropy and assuming Big Bang
nucleosynthesis value for the baryon abundance we find that only a
small fraction of cold dark matter, $\Omega_{mcp}h_0^2 < 0.007$ (at
95\%~CL), may  consists 
of millicharged
particles with the parameters (charge and mass) from this region.  
This bound significantly narrows the allowed range of the parameters of
millicharged particles. In models without paraphoton
 millicharged particles are now excluded as a dark
matter candidate.  We also speculate
that recent observation of 511~keV $\gamma$-rays from the
Galactic bulge may be an indication that a (small) fraction of CDM is
comprised of the millicharged particles. 
\end{abstract}

Search for particles carrying small but non-vanishing electric
charge 
(millicharged particles) has 
long history.
If observed, millicharged particles would either cause
serious doubts on 
the concept of Grand Unification or imply the
existence of a new massless gauge boson --- paraphoton \cite{Okun,Holdom}.
Furthermore, the existence of millicharged particles would hint towards
 processes with apparent electric charge non-conservation, like
electron or proton decay to ``nothing'' \cite{Ignatev:1978xj}.

\begin{figure}[t]
 \includegraphics[width=0.9\columnwidth,height=0.75\columnwidth]{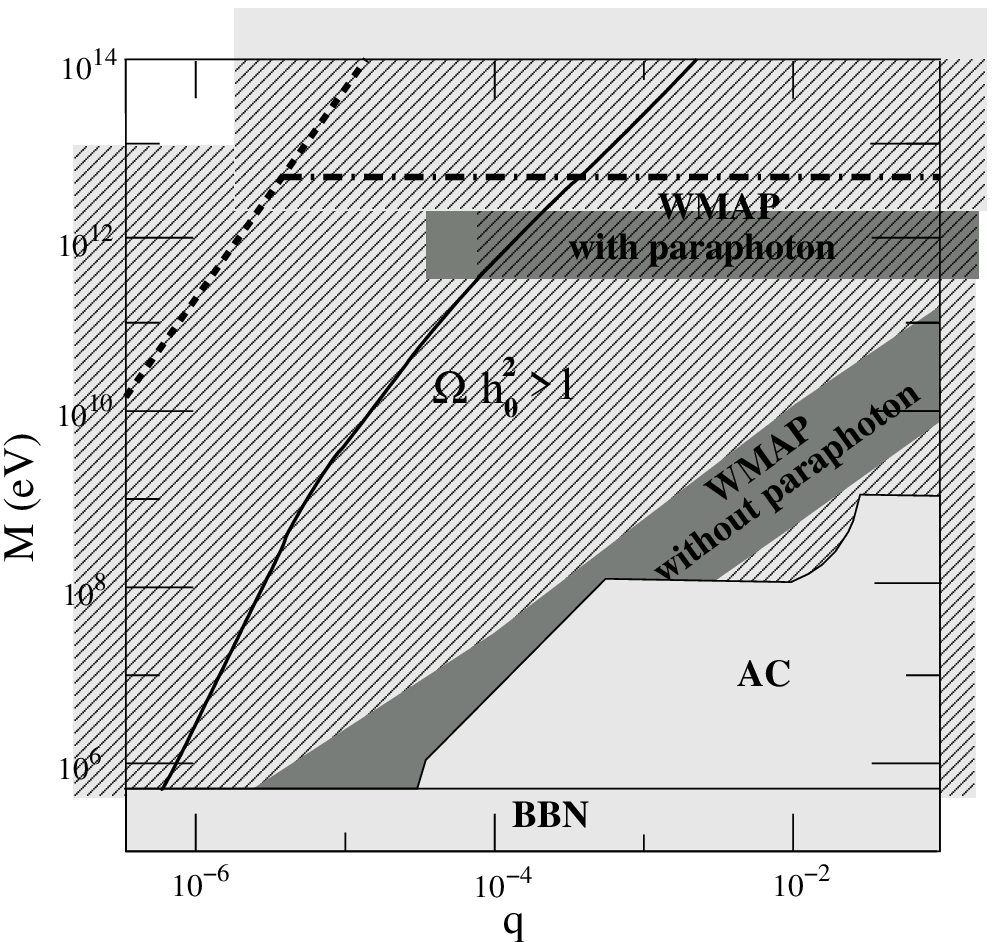} 
\caption{\label{qm} The exclusion plot in the 
parameter space for millicharged particles. Light grey area is
excluded by  accelerator experiments and  BBN. 
Dashed region is excluded by the relic abundance of
millicharged particles in models without paraphoton. 
Part of this region above the dash-dotted line is 
excluded by the relic abundance in models with paraphoton (assuming 
$\alpha'=0.1$).
On the left of the dotted
line millicharged particles cannot be thermally produced in the Early Universe.
Dark grey and dashed dark grey areas are the previously allowed
regions which are now excluded by 
Eq.~(\ref{result}) in models without and with
paraphoton, respectively. 
On the right of the  dark solid line 
millicharged particles are
coupled to baryons (see Eq.~(\ref{tightweak}) in the text).}
\end{figure}

There are various constraints on the parameters (charge and mass)
of millicharged particles, coming from collider and laboratory
experiments and from cosmology and astrophysics (see, e.g., 
Refs.~\cite{Dobroliubov:1989mr,Davidson:si,Davidson:1993sj,Prinz:1998ua,
Mohapatra:1990vq, Mohapatra:1991as}
for the latest results and Refs.~\cite{davidshannestad,Perl:2001xi}
for reviews), see Fig.~\ref{qm}.  Interestingly, reported
bounds did not
exclude a possibility~\cite{goldberghall} that a significant part (or even all) of the cold dark matter
(CDM) is comprised of millicharged particles.

The constraints on the parameters
of millicharged particles are somewhat different in theories with
and without paraphoton.
Without paraphoton, two domains in the 
parameter space of millicharged particles
are allowed.  The
first one corresponds to heavy particles with tiny
electric charge (left upper corner in Fig.~\ref{qm}) which would
never be produced thermally in the early Universe.  This region is
far beyond the reach of future collider and laboratory
experiments.  In the current {\it Letter} we are concerned with
another region. This is a narrow window of relatively light
particles with masses $M\sim 10^{-3}\div 10^2$~GeV and charges $q\sim
10^{-6}\div 10^{-1}$. Larger charges 
$q\gtrsim 0.1$ are ruled out by limits on the cosmic ray fluxes of
fractionally charged particles \cite{Perl:2001xi}. 
It is worth noting that particles with $q\gtrsim 0.2$ are also 
excluded by  measurements of the width of Z-boson 
(cf. Ref.~\cite{Davidson:si}) 
if one makes use of the latest data~\cite{PDG}.

In model with paraphoton, and for not very small values of the
 paraphoton coupling constant $\alpha'$,
 millicharged particles annihilate mainly into pairs of paraphotons.
As a result, their annihilation in the early Universe is more
 efficient
 and the cosmological bound coming from the relic abundance 
depends on the value of $\alpha'$ and generically
is less restrictive (see Fig. \ref{qm}) than in the model without
 paraphoton.

It was noted in Ref.~\cite{base}, that 
there is a part of the parameter space for the
millicharged particles where they
do not decouple from the acoustic oscillations of the
baryon-photon plasma at recombination, and it was suggested that the
effect of these 
particles on the CMB anisotropy spectrum may be similar to
the effect of baryons. 
The purpose of this {\it Letter}
is, using the recent precise CMB data from WMAP \cite{wmap},
 to set an upper limit
on the fraction of millicharged particles in CDM and to narrow 
the allowed window for millicharged particles.
Assuming the standard Big Bang nucleosynthesis (BBN) 
value for the baryon abundance, 
$\Omega_bh_0^2=0.0214\pm 0.0020$~\cite{BBN}, we arrive at 
the following constraint on the millicharged particle abundance,
\begin{equation}
\label{result}
\Omega_{mcp}h_0^2<0.007\;\; (95\%~{\rm CL})\;,
\end{equation}
if millicharged particles are coupled to baryons 
at recombination. 
The latter condition is satisfied on the right of the dark solid line
in Fig.~\ref{qm}. We see that 
the upper limit~(\ref{result}) apllies in the whole 
allowed window for millicharged
particles in models without paraphoton.
It is also worth stressing that in models with paraphoton 
the domain of applicability
of the upper bound (\ref{result}) does not depend on the value of
$\alpha'$. 
Using  the Lee-Weinberg formula
\cite{Lee:1977ua}
for the relic abundance one then translates the upper bound
(\ref{result}) into the lower limit on the annihilation cross section
of millicharged particles. 
As a result, the upper bound (\ref{result})
 excludes most part of the allowed
window for the millicharged particles in models without paraphoton, 
leaving a small allowed region
with masses in the range $m\sim 10^{-1}-10$~GeV and charges in the
range $q\sim 10^{-3}-10^{-1}$. In models with paraphoton, the bound
(\ref{result}) translates into meaningful limit on the parameters of 
millicharged particles as shown in Fig.~\ref{qm}; the excluded region
depends on the value of $\alpha'$. 

Let us proceed to the derivation of the upper bound (\ref{result}).
To calculate the CMB anisotropy spectrum we  adapt the CMBFAST
code~\cite{SeljakZaldar} which solves numerically the set of kinetic
equations \cite{mabert}
for
the linear perturbations in the primordial plasma.
To take into account the presence of 
millicharged particles we extend this set by adding the kinetic equations
for the millicharged component, modify the equations for the 
baryon component to take into account
the elastic scattering off 
millicharged particles and include millicharged component contribution
to the energy-momentum tensor. The rest of the perturbation equations are
the same as in Ref.~\cite{mabert}. 
The Compton scattering off millicharged particles is negligible, since the
corresponding cross section is suppressed by the fourth power of the
charge $q$.

We work in synchronous gauge and 
consider primordial plasma 
in the expanding Universe with  scale factor 
$a(\tau)$ (where $\tau$ is a conformal time) 
normalized to unity at present time. 
Let $T_f$,  $\rho_f$,   $\vec v_f$ be the temperature,  density and
 velocity of the $f$-th component of the plasma.
In particular,  $f=e,b,\gamma,mcp$  for electrons, baryons, 
photons and millicharged
particles, respectively. In what follows bar denotes space averaging.
The standard variables describing fluid perturbations are 
$
\delta_f(\vec k,\tau) = 
[\rho_f(\vec k,\tau) - \bar \rho_f(\tau)]/\bar\rho_f(\tau)
$
and
$
\theta_f(\vec k,\tau) = i k_j v_f^j(\vec k,\tau)
%\end{gather}
$
where $k_i$ is conformal momentum.

Before recombination, the interaction between non-relativistic electrons
and protons is strong enough to ensure that electron and baryon 
components have equal velocities, $\theta_e=\theta_b$. This makes
it possible to
use  tight coupling approximation and consider
electrons and protons as single baryon fluid~\cite{peeblesyu}. 
Then the set of equations for baryons and 
millicharged particles reads (cf. Ref.~\cite{mabert})
\begin{align}
\dot\delta_b &= - \theta_b - \frac{1}{2} \dot
h,\nonumber
\\\nonumber
\dot\theta_b &= - \frac{\dot a}{a} \theta_b + c_s^2 k^2 \delta_b 
+ \frac{4 \bar \rho_\gamma}{3 \bar \rho_b} a n_e 
\sigma_T (\theta_\gamma - \theta_b) \\\label{thetab} &+ 
a\Gamma_{mcp} \Omega_{mcp} (\theta_{mcp} - \theta_{b}), 
%\\
\end{align}
\begin{align}
\nonumber
\dot\delta_{mcp} &= - \theta_{mcp} - \frac{1}{2} \dot h,
\\
\dot\theta_{mcp} &= -\frac{\dot a}{a} \theta_{mcp} + 
c_{s,mcp}^2 k^2 \delta_{mcp} 
\nonumber \\\label{thetam} &+ a\Gamma_{mcp} \Omega_b (\theta_b - \theta_{mcp}),
\end{align}
where $h$ is the longitudinal metric perturbation, 
dot stands for derivative with respect to the conformal
time $\tau$; $c_s$, $c_{s,mcp}$ are the sound velocities in the baryon and
millicharged components, $n_e$ is the number density of electrons
and $\Gamma_{mcp}$ is the velocity transfer rate for millicharged
particles due to
scattering off baryons and electrons. The latter is given by
\begin{equation}
\label{rate}
\Gamma_{mcp}=\sum\limits_{x=e,p} \frac{n_x}{\Omega_b} \left.\frac{\partial}{\partial v_{M,x}}\left<\int \Delta v_M\cdot d\sigma_{M,x} \right>\right|_{v_{M,x}=0}   ,
\end{equation}
where brackets stand for thermal averaging, $v_{M,e(p)}$ is
relative velocity and $\Delta v_M$ is velocity transfer in a single
process of scattering; $d\sigma_{M,e}$ ($d\sigma_{M,p}$) is the
Rutherford cross section for millicharged
particles scattering off electron (proton). 

The Rutherford cross section is singular at zero scattering angle, but
due to Debye screening the integral
in Eq.~(\ref{rate}) is cut at the value of the scattering angle equal to 
the Debye angle, $\theta_D =\sqrt{2\pi\alpha n_e/T^2m_e}$. 
As a result one arrives at the following expression for the
velocity transfer rate in the case of thermal equilibrium
\begin{equation}
\Gamma_{mcp} =   \frac{4\sqrt{2\pi} 
\alpha^2 q^2 \rho_{crit}}{3 M m_p a^{3/2}(\tau)T_{0}^{3/2}} |\ln \theta_D| 
(\sqrt{\mu_{M,e}} + \sqrt{\mu_{M,p}})\;,
\label{R1}
\end{equation}
where $\mu_{M,e(p)}$ is the reduced
mass of a millicharged particle and electron (proton), 
$\alpha$ is the fine structure
constant and $T_0\approx 2.726$~K is the present CMB temperature.
It is straightforward to generalize Eq.~(\ref{R1}) 
to nonequilibrium case when electrons, protons and millicharged
particles have different temperatures. In that case
 the value of $\Gamma_{mcp}$  is larger than the one
given by Eq.~(\ref{R1}), so Eq.~(\ref{R1}) 
may be used as a lower estimate of the interaction rate, which is
sufficient for our purposes.
 
We solve the system of the kinetic equations starting from 
the early moment of time $\tau_i$ and using
inflationary initial conditions~\cite{mabert},
\begin{align}\nonumber
\delta_{mcp}& = \delta_{b} = \frac{3}{4} \delta_\gamma = -\frac{1}{2} C (k \tau_i)^2,
\\\nonumber
\theta_{mcp}& = \theta_{\gamma} = \theta_{b} = - \frac{1}{18} Ck (k\tau_i)^3\;,
\end{align}
where  constant $C$ determines the overall
normalization.

The last terms in the r.h.s. of Eqs.~(\ref{thetab}), 
(\ref{thetam}) tend to equalize the velocities of
the baryon and
millicharged components of the fluid. 
This results in the kinetic
equilibrium, $\theta_{mcp} = \theta_b$, provided 
$\Gamma_{mcp}$ is large enough. 
In this case,  the
perturbation equations are difficult to solve numerically,
because the kinetic relaxation rate for the interaction of millicharged
particles and baryons is much larger than the rates of other processes.
To deal with this situation we make use of the zeroth order
tight coupling approximation adopting the method of Peebles and 
Yu~\cite{peeblesyu}. 
Namely, we expand Eqs.~(\ref{thetab}),~(\ref{thetam}) to
the zeroth order in  $\Gamma_{mcp}^{-1}$, setting 
$\theta_b = \theta_{mcp} = \theta$. 
Then we exclude $\Gamma_{mcp}$ from Eqs.~(\ref{thetab}),
(\ref{thetam})
and arrive at the following equation
\begin{equation}
\label{tight0}
\begin{aligned}
\dot\theta = 
&
- \frac{\dot a}{a} \theta +
\frac{\Omega_b}{\Omega_b + 
\Omega_{mcp}} \frac{4 \bar \rho_\gamma}{3 \bar \rho_b} 
a n_e \sigma_T (\theta_\gamma - \theta)
\\ &
+ \frac{\Omega_b~ c_s^2 \delta_b + 
\Omega_{mcp}~c_{s,mcp}^2 \delta_{mcp}}{\Omega_b+\Omega_{mcp}}k^2\;.
\end{aligned}
\end{equation}
The CMB spectrum obtained 
in this approximation agrees with the solution of the
original set of equations at the level of one percent provided that 
\begin{equation}
\label{tight}
\Gamma_{mcp}(\tau_{rec}) (\Omega_b +
 \Omega_{mcp}) H(\tau_{rec})^{-1} \gtrsim 250\;,
\end{equation}
where $\tau_{rec}$ is conformal time at recombination
and 
$H(\tau_{rec})$ is the Hubble parameter. 
In what follows we discuss a region of parameters where 
tight coupling condition (\ref{tight}) is
 satisfied (in particular, this region covers the whole allowed window
 in models without paraphoton)
 and comment on 
the rest of the parameter space in due course. 

To compare the results of our simulation with the CMB data
we consider the flat $\Lambda$CDM model with the number of massless
neutrino species $N_{\nu}=3$.
We perform a scan over the space of the cosmological models by varying
parameters from minimal to
maximal values as given in Table~\ref{matr}.
All priors  are  at 95\%~CL. 
 \begin{table}[htb]
\begin{tabular}{| c| c| c| c| c|}
\hline $parameter$& $min.~value$ & $max.~value$ & $step$ & reference \\
\hline $\Omega_{ C \! D \! M}$ & 0.2 & 0.4 & 0.01 & PDG~\cite{PDG}\\\hline
$h_0$ & 0.64 & 0.79 & 0.01 & PDG~\cite{PDG}\\\hline
$n_s$ & 0.8 & 1.2 & 0.01 & \\\hline
$\Omega_b h_0^2$ & 0.0194 & 0.0234 & 0.0005 & BBN~\cite{BBN} \\\hline
$\Omega_{mcp}$& 0 & 0.020 & 0.001 &\\\hline
\end{tabular}
\caption{\label{matr} The ranges of the cosmological parameters used
in simulations; we never have a good fit
outside the region $n_s \in [0.8, 1.2]$, and for $\Omega_{mcp}>0.02$.}
\end{table}
We assume helium
fraction at the moment of recombination $Y_{He}=0.24$. We have checked
that reionization effects are irrelevant here, the reason being that
 reionization affects the spectrum only at the lowest
values of the 
multipole moments.  

For each cosmological model we calculated likelihood to the WMAP
data~\cite{wmap}. 
As a result we arrived at the limit given by Eq.~(\ref{result}),
which means that no model exists in the considered parameter range with
larger values of $\Omega_{mcp}h_0^2$ and likelihood better than 5\%.  
We  also checked that additional CMB data 
from CBI~\cite{CBI} and
ACBAR~\cite{ACBAR} do not improve the limit~(\ref{result}). 

Using  the Lee--Weinberg formula we translate the limit given by
Eq.~(\ref{result})
into the bound on the parameter space. The corresponding 
excluded areas 
are shown in Fig.~\ref{qm}. In models with paraphoton we extend our
analysis 
to the region where the tight coupling condition (\ref{tight}) is not 
satisfied. We  checked that actually
the upper bound (\ref{result}) applies provided that the following
less restrictive condition holds 
\begin{equation}
\label{tightweak}
\Gamma_{mcp}(\tau_{rec}) (\Omega_b +
 \Omega_{mcp}) H(\tau_{rec})^{-1} \gtrsim 2.5\;. 
\end{equation} 
This inequality is true on the right 
of the black solid line in Fig.~\ref{qm}. 

\begin{figure}[t]
 \includegraphics[width=0.9\columnwidth]{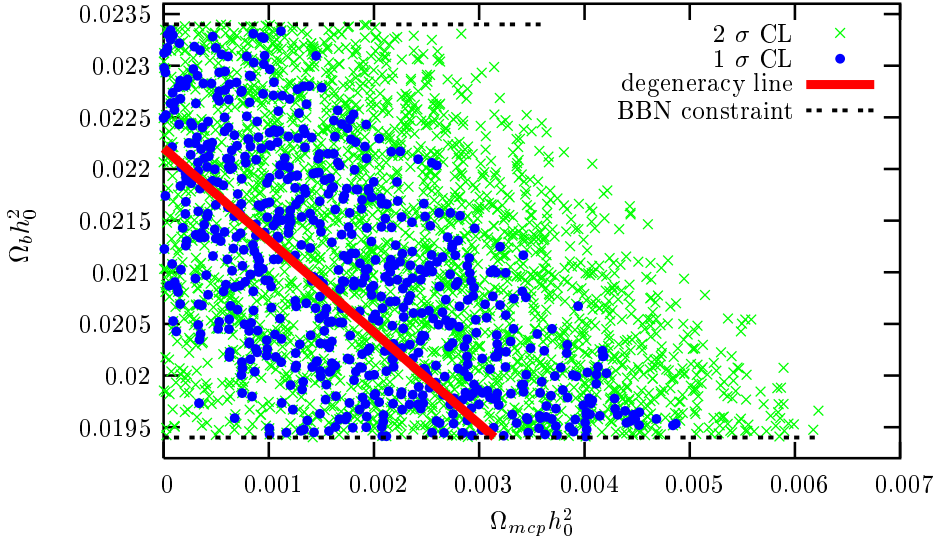} 
\caption{\label{distr} Distribution of
models in the ($\Omega_b
h_0^2$, $\Omega_{mcp} h_0^2$) plane. Crosses and dots
denote models agreeing
with data at the
2$\sigma$ and 1$\sigma$ CL, respectively. The bold line illustrates the degeneracy
of the CMB anisotropy spectrum. Two dotted lines show the range of
$\Omega_b h_0^2$ allowed by  BBN.}
\end{figure}

Coming back to the tight coupling regime, let us
illustrate the upper bound (\ref{result})
 pictorially by generating a set of 
models with random parameters in tight coupling regime, assuming uniform 
distribution of models in the parameter space within ranges given in
Table~\ref{matr}. For each model we calculated likelihood 
to the observed CMB spectrum and
plot 
the resulting distribution of models in the 
($\Omega_b h_0^2$,~$\Omega_{mcp} h_0^2$) plane in Fig.~\ref{distr}.
 One observes that the CMB spectrum is approximately degenerate
along the straight line 
\begin{equation}
\label{deg}
\Omega_b h_0^2 = 0.022 - 1.1~ \Omega_{mcp} h_0^2.
\end{equation} 
This degeneracy is in agreement with the expectation of
Ref.~\cite{base},
that the effect of millicharged particles is similar to that of
baryons.
Thus in models with millicharged particles, the CMB data~\cite{wmap}
determine actually the sum $(\Omega_b+\Omega_{mcp}) h_0^2=0.022\pm
0.001$ (68\% CL). Combining this value
with the lower limit $\Omega_b h_0^2>0.019$ from  BBN
one arrives at the upper bound very
similar to Eq. (\ref{result}). This serves as a qualitative
explanation of our result.

\begin{figure}[t]
\includegraphics[width=0.9\columnwidth]{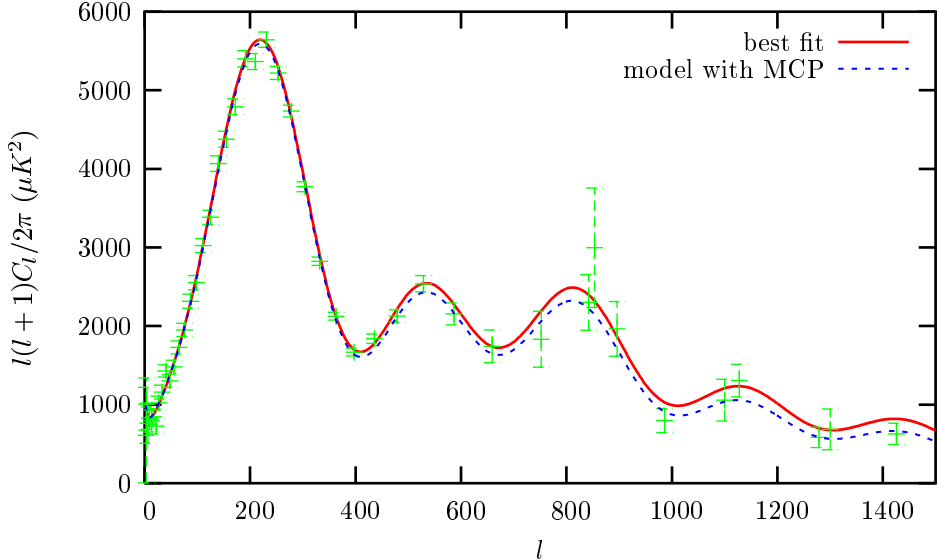} 
\caption{\label{cmbspec}Two different CMB anisotropy spectra 
compared with extended WMAP dataset. Solid line represents the best fit
model without millicharged particles, $\Omega_bh_0^2 = 0.022$. 
Dashed line 
corresponds to model with $\Omega_b h_0^2= 0.014, \Omega_{mcp}h_0^2 =
0.007$.}
\end{figure}

Another illustration of the approximate degeneracy (\ref{deg})
is shown in Fig.~\ref{cmbspec},  
where two CMB anisotropy spectra calculated for
different models on the degeneracy line are shown. 
One observes that the two spectra
almost coincide in the region of the first and second acoustic peaks.
However, the degeneracy is no longer present 
at higher multipoles.
This is due to the fact that the electroneutrality of the
plasma implies that the electron number density is proportional 
to the baryon
density. Hence, replacing a certain amount of baryons by
millicharged particles results in the enhancement of the Silk damping at
small scales.
With future precise data for high values of $l$,
one will be able to set a constraint on the value of 
$\Omega_{mcp}$ using the CMB data only,
without  reference to BBN results. To check this
 we created a simulated dataset, which contains the same
values of $l$ as in the WMAP data up to $l=500$, and then with the step
$\Delta l=50$ up
to $l=1600$. The CMB anisotropy spectral coefficients $C_l$'s were
taken from the best fit~\cite{wmap} to WMAP data. 
The error bars for these coefficients were assumed to be equal to
cosmic variance. Repeating the above procedure for 
this dataset we obtained that an upper limit
$\Omega_{mcp}h_0^2<0.003$ can be placed. 
Further improvement of this limit turns out
to be impossible due to the new approximate degeneracy,
\begin{align}\nonumber
\Omega_b h_0^2  = 0.022 - 0.65~ \Omega_{mcp} h_0^2\;,\\
n_s =0.94+8.0~\Omega_{mcp}\;,\label{deg2}
\end{align}
arising at smaller values of $\Omega_{mcp}h_0^2$.

To conclude, we note that when translated into the parameter space, the
limit (\ref{result}) is especially interesting for the models without paraphoton,
where it excludes most part of the window with not very heavy
particles and substantial electric charges. To completely close
the window,  sensitivity
to millicharged particle abundance at the level of 
$\Omega_{mcp}h_0^2\sim 3\cdot 10^{-4}$ would be required,
which cannot be achieved with
future CMB experiments due to the degeneracy 
(\ref{deg2}). Determination of the baryon
abundance from the BBN is not accurate enough to improve the
situation. Hopefully, the rest of the window will be explored by future
accelerator and/or laboratory experiments.

Finally, recently it was suggested \cite{Boehm:2003bt} that
the flux of the 511 keV $\gamma$-rays from the Galactic bulge detected
by the INTEGRAL satellite \cite{Knodlseder:2003sv}
may be explained by the annihilation of the $\sim 1\div 100$ MeV dark
matter particles into $e^+e^-$ pairs, provided their 
annihilation cross section
$\sigma\beta$ and abundance $\Omega$ satisfy
$
\left(
\frac{\sigma\beta}{\rm{pb}}\right)
\left(\frac{1\rm{MeV}}{M}\right)^2\left(
\frac{\Omega^2}{\Omega^2_{C \! D \! M}}\right)\simeq 
10^{-(3.5\div 4.5)}
$. 
Intriguingly, this condition holds in the left corner
of the parameter space for millicharged particles without paraphoton
allowed by
Eq. (\ref{result}) (say, for $q=3\cdot 10^{-3}$, $M=100$ MeV).
One is tempted to speculate that the observation of the 511 keV
line is an indication that a (small) fraction of CDM 
is comprised of the millicharged particles. This possibility will be
 checked by the future CMB data.
\newline
{\it Acknowledgments.} We would like to thank V.Rubakov and
S.Sibiryakov for useful discussions. This work was supported in part
by the RFBR02-02-17398 and GPRFNS-2184.2003.2 grants. 
The work of D.G. was also supported in part by the
GPRF grant
MK-2788.2003.02.
\bibliographystyle{amsplain}

\end{document}